\documentclass[pra,reprint]{revtex4-1}
\usepackage{graphicx}
\usepackage{hyperref}
\usepackage{verbatim}
\usepackage{bbold,bbm}
\usepackage{xcolor}
\usepackage{amsmath,amssymb}
\usepackage{listings}
\begin{document}

\title{Graphical Calculus for products and convolutions}
\author{Filippo M. Miatto}
\affiliation{Institut Polytechnique de Paris}
\affiliation{T\'el\'ecom ParisTech, LTCI, 46 Rue Barrault, 75013 Paris, France}

\begin{abstract}
Graphical calculus is an intuitive visual notation for manipulating tensors and index contractions. Using graphical calculus leads to simple and memorable derivations, and with a bit of practice one can learn to prove complex identities even without the need for pen and paper. This manuscript is meant as a demonstration of the power and flexibility of graphical notation and we advocate exploring the use of graphical calculus in undergraduate courses.
In the first part we define the following matrix products in graphical language: dot, tensor, Kronecker, Hadamard, Kathri-Rao and Tracy-Singh. We then use our definitions to prove several known identities in an entirely graphical way. Despite ordinary proofs consist in several lines of quite involved mathematical expressions, graphical calculus is so expressive that after writing an identity in graphical form we can realise by visual inspection that it is in fact true. As an example of the intuitiveness of graphical proofs, we derive two new identities.
In the second part we develop a graphical description of convolutions, which is a central ingredient of convolutional neural networks and signal processing. Our single definition includes as special cases the circular discrete convolution and the cross-correlation. We illustrate how convolution can be seen as another type of product and we derive a generalised convolution theorem. We conclude with a quick guide on implementing tensor contractions in python.

\end{abstract}
\maketitle

\section{Introduction}
When we manipulate mathematical expressions (either in writing or mentally), the behaviour of the typographical symbols can acquire its own physicality. For example, we might feel that as soon as we wrap a product of various terms in a logarithm, the ``force'' that is binding them together comes loose, as $\log(xyz) = \log(x)+\log(y)+\log(z)$. Even the simple operation of bringing the denominator of one side of an equation to the numerator of the opposite side ($\frac{x}{2}=y\ \rightarrow\ x=2y$) is a physical interpretation of the actual operation of multiplying both sides by the same quantity.
Such physical familiarity with symbolic manipulation comes with time and practice, and arguably the more a notion is described by symbols that behave somewhat physically, the easier it is to have such an experience. Graphical calculus is an extreme example of such notation: in graphical calculus all the operations consist in connecting ``bendable'' and ``stretchable'' wires, and one can perform complicated tensor manipulations entirely within this framework.

Graphical calculus offers a representation of concepts that complements the standard mathematical notation: it is by providing alternative representations of the same concept that we help the recognition networks in the brain of a learner, as recommended in the principles of universal design for learning \cite{rose2000universal}. It is with this student-centered spirit that we present this paper.


In the first part of this work we define six types of matrix product (dot, tensor, Hadamard, Kronecker, Kathri-Rao, Tracy-Singh) in graphical calculus notation. Our aim is to make them available in a form that is easy to remember, easy to work with and easy to explain to others. To stress the power of graphical calculus, we show two new (to the best of the author's knowledge) identities. In the second part we describe a generalized discrete convolution which includes the standard convolution and the cross-correlation as special cases. These operations are extremely common in signal processing and in the field of machine learning, in particular in convolutional neural networks \cite{lecun1989backpropagation} and signal processing.  

Graphical calculus was introduced in the seventies by Penrose \cite{penrose1971applications} in the context of general relativity, it was then slowly picked up and/or rediscovered by various authors including Lafont \cite{lafont2003towards}, Coecke and Abramsky \cite{abramsky2004categorical, coecke2010quantum}, Griffiths \cite{griffiths2006atemporal}, Seilinger \cite{selinger2010survey}, Baez \cite{baez2010physics}, Wood \cite{wood2011tensor}, Biamonte \cite{biamonte2017charged}, Jaffe \cite{jaffe2018holographic} and others \cite{backens2014zx, jeandel2018complete}. Because of such diversity of scopes and the scarcity of mainstream adoption, the notation is not yet standardised. Some authors proceed vertically, others horizontally. Some right to left, others left to right. Obviously, there is no \emph{actual} difference, but different notations can make it easier or harder to transition between regular and graphical notations. In our case, we opted for a horizontal notation (as it is more similar to the way in which we write) and right to left (as that is how we compose successive matrix multiplications).

A few words about our conventions. When we write tensors with indices (usually roman literals), we use the convention that repeated indices are implicitly summed over, to avoid an excessive use of the summation symbol. When we contract a pair of indices we assume that their dimensions match. In graphical notation, a tensor of rank $R$ is drawn as a shape with $R$ wires, where each wire represents an index. When we connect two wires it means we are summing over those indices, like so: $A_{ij}$ and $B_{jk}$ both have two indices (i.e. we can think of $A$ and $B$ as matrices) which means that in graphical notation they both have two wires. If we join the two wires corresponding to the $j$ indices (assuming they have the same dimension), we obtain the tensor $A_{ij}B_{jk}$ which has two leftover indices because we sum over the repeated $j$ index: $\sum_j A_{ij}B_{jk}$ (see Fig.~\ref{dotprod}). Technically it doesn't matter how we orient the drawings of the tensors and their wires, as long as we keep track of which index they correspond to. However, in order to help  transition to regular notation, we orient row wires toward the left and column wires toward the right.

\section{Two mediating tensors}
In order to compute the various products that we are going do describe, we will need the help of the Kronecker and the vectorization tensors, which are two ``mediating'' tensors that enable such products. In the following subsections we will see them separately and get a feeling for what they do, both at the level of indexed notation and at the level of graphical notation. 

\subsection{The Kronecker tensor}
\begin{figure}[h!]
\centering
\includegraphics[page=1, width = \columnwidth]{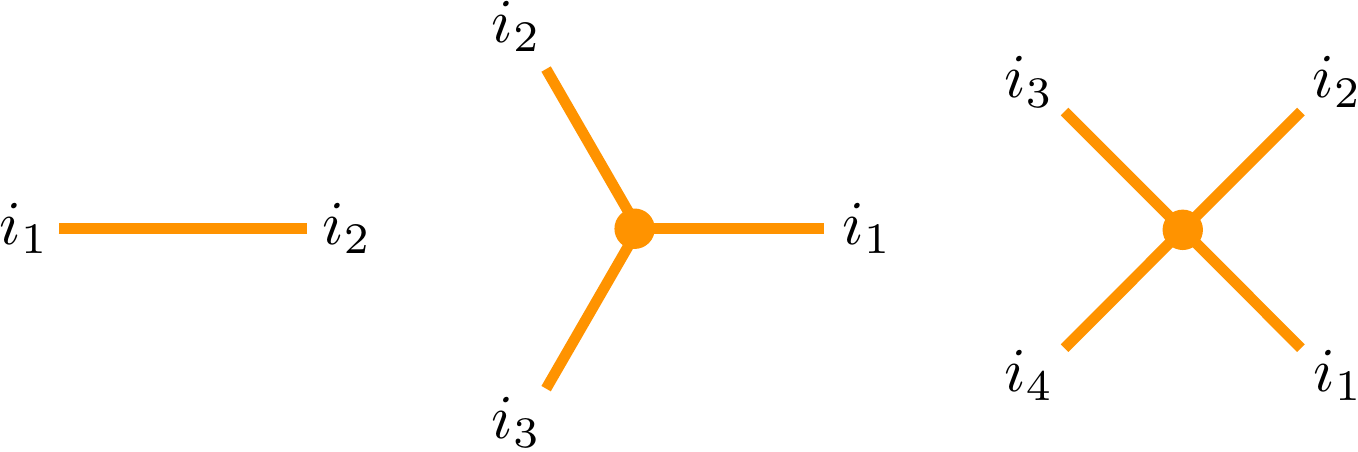}
\caption{\label{KT}The Kronecker tensor is a rank-N Kronecker delta, which is 1 only if all of the indices have the same value and 0 otherwise. Here the rank-2, rank-3 and rank-4 examples.}
\end{figure}

The $D$-dimensional, rank-$N$ Kronecker tensor is a generalization of the familiar Kronecker delta $\delta_{ij}$. It is defined as:
\begin{align}
\delta_{i_1i_2\dots i_N} = \begin{cases}1\quad \mathrm{if}\ i_1=i_2=\dots=i_N\\0\quad \mathrm{otherwise}\end{cases}
\end{align}
where each index has values in $0\dots D-1$.
As special cases, $\delta_{ij}$ is equivalent to the identity matrix and it is also equivalent to an (unnormalized) Bell state, well known in quantum information \cite{wood2011tensor}. Similarly, $\delta_{ijk}$ is equivalent to an (unnormalized) GHZ state \cite{biamonte2017charged}. In graphical notation we indicate such tensors as a small black circle with $N$ wires. In case we have only two wires we can omit the black circle (see Fig.~\ref{KT}).
Here are a few properties of the Kronecker tensor. Every $D$-dimensional Kronecker tensor, independently of its rank (i.e.~the number of wires) contains $D$ ones and the rest of its values are zeros. This is true also for the rank-1 Kronecker tensor $\delta_{i}$, which therefore is the vector with all ones. Similarly, the tensor product $\delta_{i_1}\delta_{i_2}\dots\delta_{i_N}$ is the constant tensor whose all entries are ones (in this case we can have more than $D$ ones, because the tensor product of deltas is not a delta).
\begin{figure}[h!]
\centering
\includegraphics[page=2, width = 0.5\columnwidth]{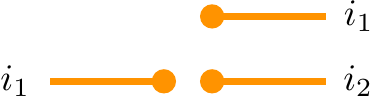}
\caption{\label{KT2}We can use multiple copies of the rank-1 Kronecker tensor to construct constant tensors whose entries have all value 1. In this figure we have a constant vector with all ones (on top) and a constant matrix with all ones (on the bottom).}
\end{figure}
The contraction of any compatible (in terms of index dimensions) Kronecker tensors yields other Kronecker tensors, e.g.~$\delta_{i_1i_2i_3i_4}\delta_{i_2i_3} \delta_{i_4}= \delta_{i_1}$. The Kronecker tensor can mediate a dot product: $A_{ij}B_{kl}$ is a tensor with four indices, but $A_{ij}B_{kl}\delta_{jk}$ is equivalent to the dot product of the two matrices $A$ and $B$. We can also generalise it to act on more than two matrices (or tensors), such as $A_{ij}B_{kl}C_{mn}\delta_{ikm}$. The Kronecker tensor can extract the diagonal of a matrix: $\mathrm{diag}(A)_k=A_{ij}\delta_{ijk}$, create a diagonal matrix from a vector: $\mathrm{diag}(\mathbf{a})_{jk} = a_i\delta_{ijk}$ or erase all of the off-diagonal elements of a matrix: $A_{ij}\delta_{ijkl}$. The Kronecker tensor can be used to compute the trace of a matrix: $\mathrm{Tr}(A)=A_{ij}\delta_{ij}$, or the partial trace of a higher order tensor: $\mathrm{Tr_1}(A)_{kl} = A_{ijkl}\delta_{ij}$, which is especially useful in quantum information.
\begin{figure}[h!]
\centering
\includegraphics[page=3, width = \columnwidth]{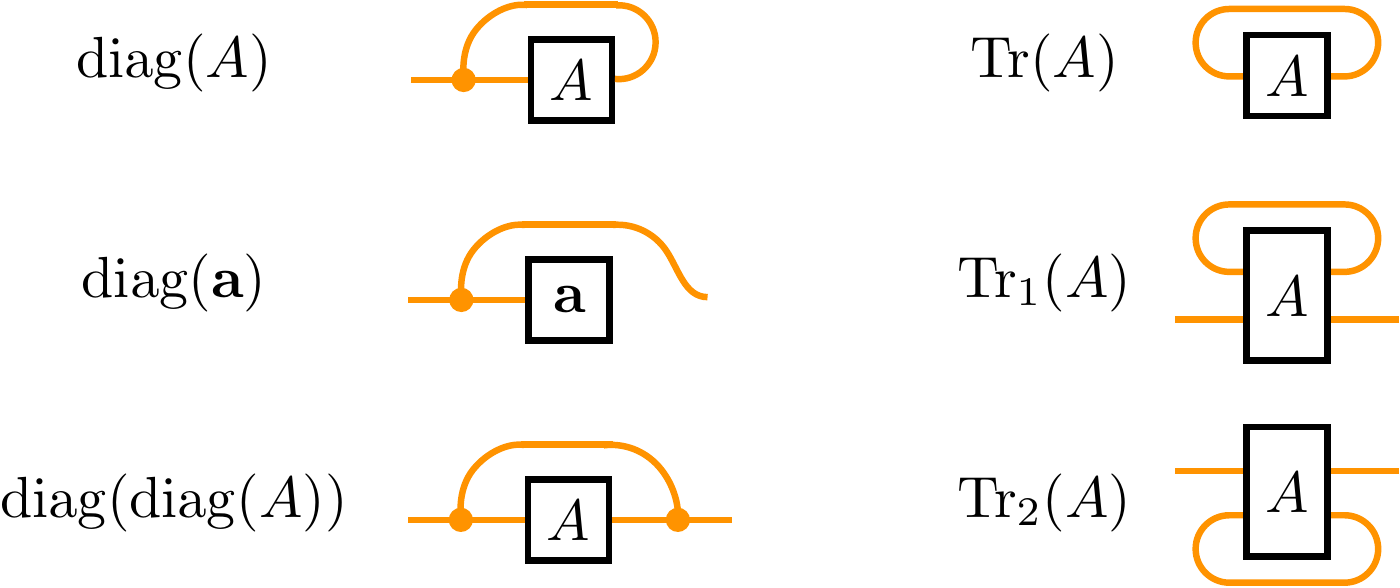}
\caption{\label{KTp}The Kronecker tensor can extract the diagonal of a matrix $A$, produce a diagonal matrix from a vector $\mathbf{a}$, erase the off-diagonal elements of $A$, compute traces and partial traces. Note that we are indicating two operations with the same name $\mathrm{diag}(\cdot)$ as their meaning is clear from the context (diag of a matrix yields a vector, and diag of a vector yields a matrix).}
\end{figure}

\subsection{The vectorization tensor}
The vectorization tensor mediates the operation of ``serializing'' multiple indices into one. For example, if we have two indices $i\in\{0,\dots,I-1\}$ and $j\in\{0,\dots,J-1\}$, we can combine them into a single index $m = i+jI$ with $m\in\{0,\dots,IJ-1\}$, and we can always retrieve $i$ and $j$ from $m$ if we know $I$ and $J$. What happens to a matrix $A_{ij}$ that is vectorized is that its columns (or rows, depending on how we choose to perform the contraction) are concatenated in a vector so that all of the entries are now indexed by $m$. The generalized index formula for $N$ indices $i_1, i_2, \dots, i_N$ with values in the ranges $I_1, I_2,\dots,I_N$ is
\begin{align}
m &= \sum_{j=1}^Ni_j\left(\prod_{\ell=1}^{j-1}I_\ell\right) \\&= i_1+i_2 I_1+i_3 I_1I_2+\dots+i_NI_1I_2\dots I_{N-1}
\end{align}
We call it vectorization tensor because it flattens all of the wires that it is attached to into a single one (like a vector). We indicate it as $\gamma_{i_1i_2\dots,m}$, notice the comma between the indices being vectorized and the new index. The indexed definition of the vectorization tensor is
\begin{align}
\gamma_{i_1i_2\dots i_N,m} = \begin{cases}1\quad \mathrm{if}\ m = \sum_{j=1}^Ni_j\left(\prod_{\ell=1}^{j-1}I_\ell\right)\\0\quad \mathrm{otherwise}\end{cases}
\end{align}
We draw it as a triangle with N+1 wires:
\begin{figure}[h!]
\centering
\includegraphics[page=4, width = 0.5\columnwidth]{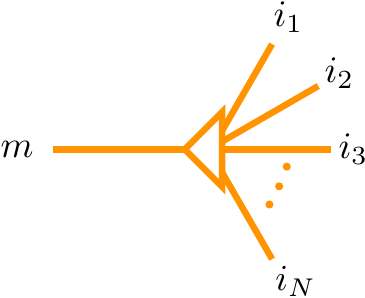}
\caption{The vectorization tensor serializes multiple indices into a single index in a reversible way (if we keep track of the dimension and position of each wire).}
\end{figure}

Note that as the vectorization tensor is not symmetric (as opposed to the Kronecker tensor), the order of the wires is crucial.

These two tensors enjoy of the following properties:
\begin{figure}[h!]
\centering
\includegraphics[page=5, width = \columnwidth]{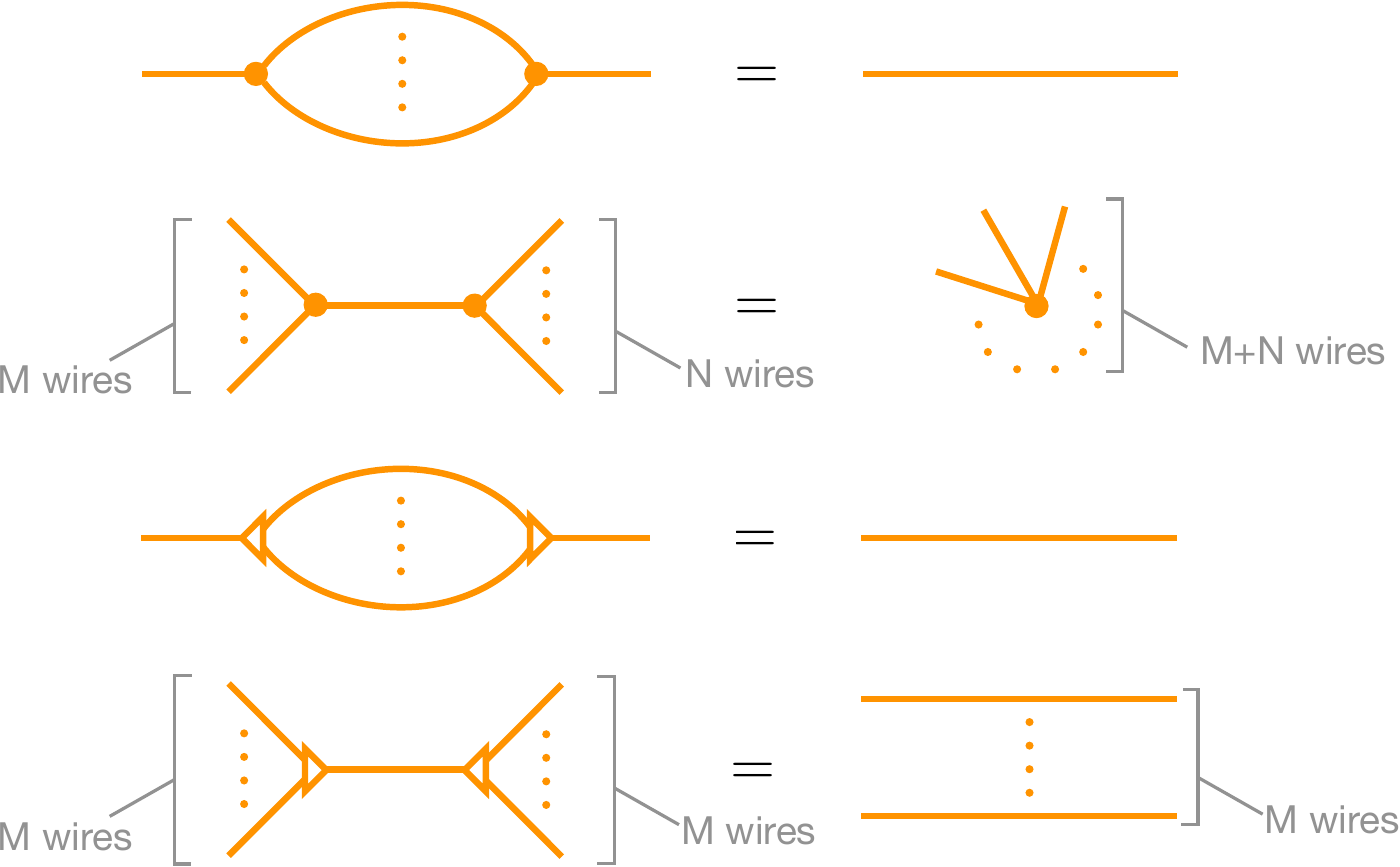}
\caption{\label{fundamentalidentities} Above: when we contract multiple Kronecker tensors with wires of the same dimension we obtain other Kronecker tensors with the appropriate number of wires. Below: when we contract vectorization tensors, we need to be careful to match the dimensions. Depending on which wires we contract, we can end up with a high-dimensional identity matrix or with the tensor product of several identity matrices.}
\end{figure}

\begin{figure}[h!]
\centering
\includegraphics[page=6, width = \columnwidth]{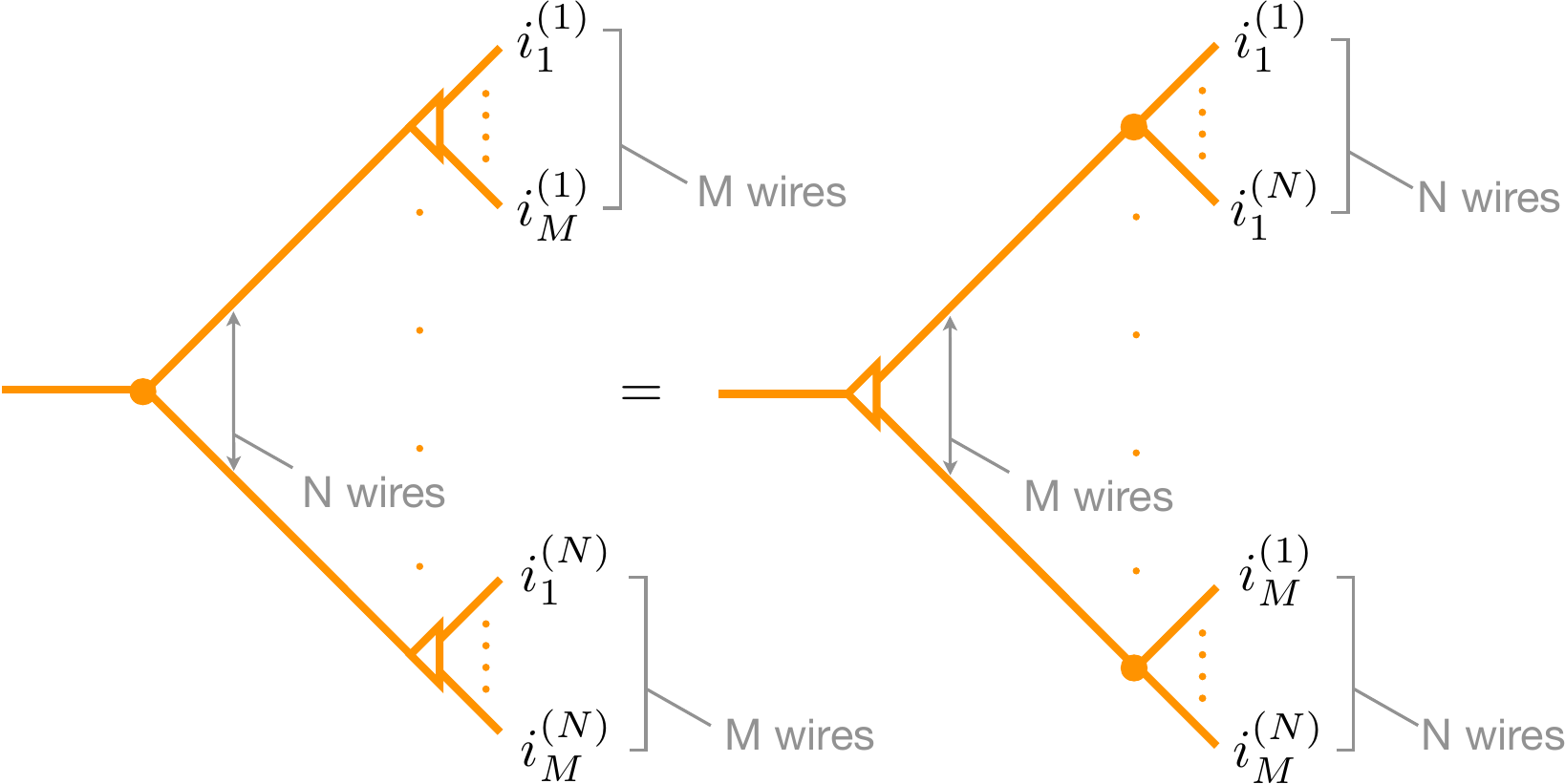}
\caption{\label{KVswap} This identity holds if in the LHS there are $N$ copies of the same vectorization tensor and the Kronecker tensor's indices have a consistent dimension. Notice that we are ``transposing'' the indices.}
\end{figure}

\begin{figure}[h!]
\centering
\includegraphics[page=7, width = 0.7\columnwidth]{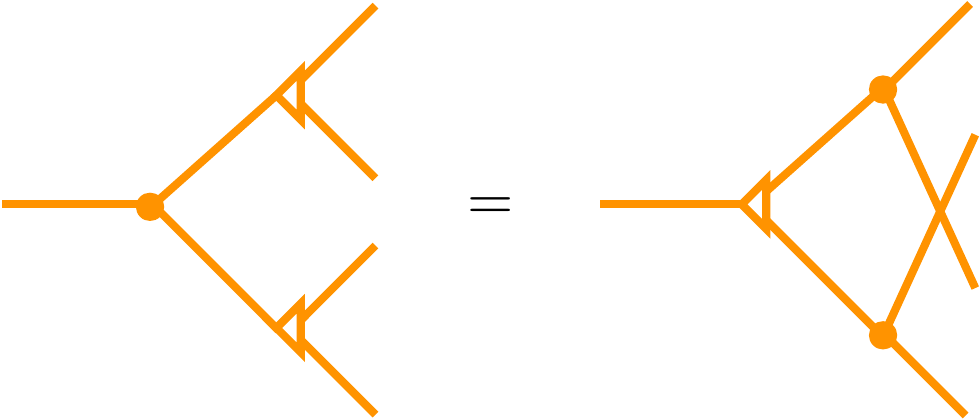}
\caption{\label{KVswapExample}As an example of the swapping rule in Fig.~\ref{KVswap}, notice how swapping the Kronecker and vectorization tensors leads to twisting wires.}
\end{figure}

The Kronecker and vectorization tensors can be composed together and if the dimensions of the wires are compatible, they enjoy of the property in Fig.~\ref{KVswap}. See also Fig.~\ref{KVswapExample} for a an example of a special case.

As we will see below, this property is the origin of the bisymmetry property of some of the products that we are going to describe.

\section{Products}
We now proceed with the presentation of a few matrix products. Here we summarize the requirements for the dimensions of the matrices/tensors in order for the products to be well-defined:
\ \\
\ \\
\ 
\begin{tabular}{l | c | c | c}
\ &$\mathrm{dim}(A)$&$\mathrm{dim}(B)$&result\\
\hline
Dot&$I\times J$&$J\times K$&$I\times K$\\
Tensor&$I\times J$&$K\times L$&$I\!\times\! J\! \times\!K\! \times\! L$\\
Kronecker&$I\times J$&$K\times L$&$IK\times JL$\\
Hadamard&$I\times J$&$I\times J$&$I\times J$\\
Kathri-Rao&$I\times J$&$K\times J$&$IK\times J$\\
Tracy-Singh&$I\!\times\! J\!\times \!K\!\times \!L$&$P\!\times \!Q\!\times \!R\!\times \!S$&$IPKR\!\times\! JQLS$\\
\end{tabular}

\subsection{Dot product}
The dot product consists in summing over a repeated row-by-column pair. The dot product described with indices is $(AB)_{ik}=A_{ij}B_{jk}$. In graphical notation we indicate it by connecting the wires corresponding to the index $j$:
\begin{figure}[h!]
\centering
\includegraphics[page=8, width = 0.4\columnwidth]{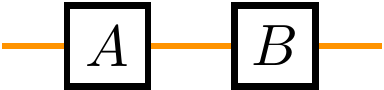}
\caption{\label{dotprod}The dot product in graphical notation consists of connecting wires directly. Each dot product reduces by 2 the total number of initial wires.}
\end{figure}\ \\
As two free wires remain, the result is another matrix.

\subsection{Tensor product}
The tensor product preserves all of the index information: $(A\otimes B)_{ijk\ell} = A_{ij}B_{k\ell}$. To perform the tensor product in graphical calculus, we simply stack the tensors and we leave their wires untouched:
\begin{figure}[h!]
\centering
\includegraphics[page=9, width = 0.25\columnwidth]{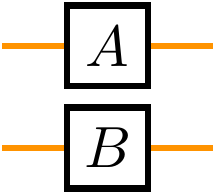}
\caption{To take the tensor product of two matrices $A$ and $B$ we stack them and preserve all of the indices. The tensor product does not change the total number of wires.}
\end{figure}

\subsection{Kronecker product}
The Kronecker product is what is often conflated with the tensor product, especially in the quantum information literature. Perhaps the fact that they are usually both indicated  with the simbol $\otimes$ contributes to the mixup. To compute the Kronecker product one begins with a tensor product, but then one makes the extra step of vectorizing groups of indices together. For example, the Kronecker product of two matrices is another matrix: $(A\otimes B)_{mn} = A_{ij}B_{k\ell}\gamma_{ik,m}\gamma_{j\ell,n}$, which means that $m\in\{0,\dots,IK-1\}$ and  $n\in\{0,\dots,JL-1\}$.
\begin{figure}[h!]
\centering
\includegraphics[page=10, width = 0.5\columnwidth]{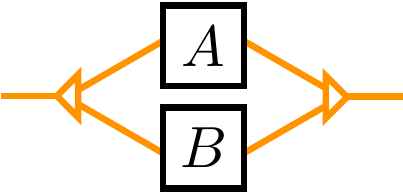}
\caption{The Kronecker product is like a tensor product, except we vectorize the row indices and the column indices. The Kronecker product reduces the total number of wires, but it is reversible, so no information is lost (unlike the Hadamard product, below).}
\end{figure}
Despite having only two wires and not four, the result contains all of the information contained in the tensor product, except for the information about which subsystem has which dimensionality. If we work with systems of equal dimension (such as a set of qubits) this is not a problem.

\subsection{Hadamard product}
The Hadamard product is an element-wise multiplication and it is defined if the two matrices match both of their dimensions exactly: $(A\circ B)_{ij}=A_{ij}B_{ij}$
\begin{figure}[h!]
\centering
\includegraphics[page=11, width = 0.5\columnwidth]{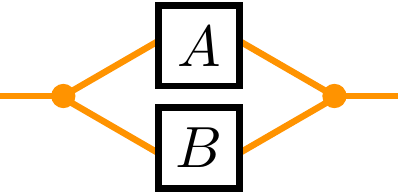}
\caption{The Hadamard product is an element-wise product, so we need the two row indices to be the same and the two column indices to be the same, which we achieve by using the Kronecker tensor.}
\end{figure}\ \\
This is clear also from the fact that we define it in terms of the Kronecker tensor, whose wires all have the same dimension. Note that we presented it for a pair of matrices, but obviously the element-wise product is defined for tensors of any rank.

\subsection{Kathri-Rao product}
This Khatri-Rao product \cite{khatri1968solutions} is useful in data processing and in optimizing the solution of inverse problems that deal with a diagonal matrix \cite{zhang2002inequalities}. The definition of the Kathri-Rao product is somewhat cryptic, but in terms of graphical calculus it is very simple. In the standard description, we have two matrices of dimension $I\times J$ and $K\times J$ therefore with the same number of columns, $A=[\mathbf{a}_0,\dots,\mathbf{a}_{J-1}]$ and $B=[\mathbf{b}_0,\dots,\mathbf{b}_{J-1}]$. The Kathri-Rao product is  the matrix with $J$ columns whose $j$-th column is the Kronecker product of columns $\mathbf{a}_j$ and $\mathbf{b}_j$: $(A\odot B) = [\mathbf{a}_0\otimes\mathbf{b}_0,\dots,\mathbf{a}_{J-1}\otimes\mathbf{b}_{J-1}]$.

\begin{figure}[h!]
\centering
\includegraphics[page=18, width = \columnwidth]{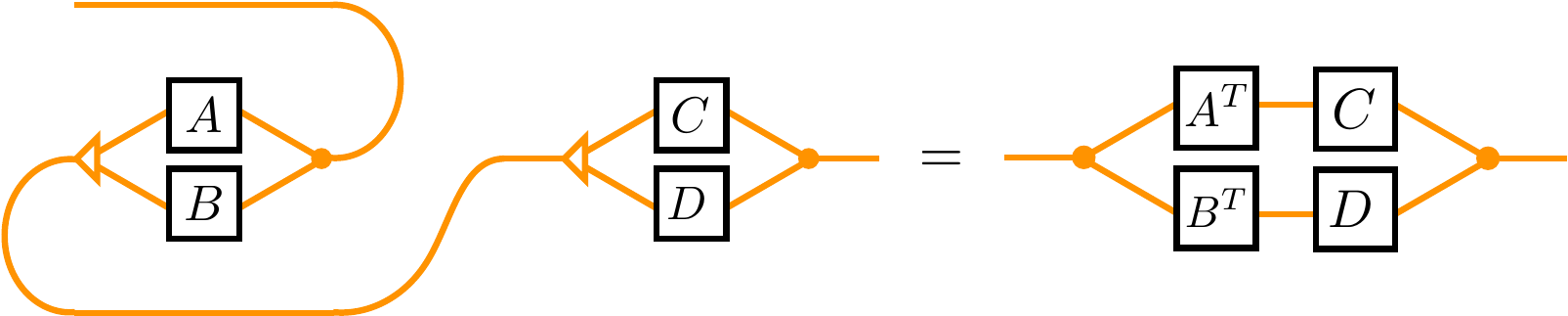}
\caption{\label{KR1} In this figure we prove the transpose property of the Kathri-Rao product: $(A\odot B)^T(C\odot D) = A^TC\circ B^TD$. This is not the bisymmetry property because the Kathri-Rao product does not appear in the right hand side, but rather the Hadamard product. Note that in the first part of the LHS we have swapped row and column indices to express the transpose. Then we have used the identity at the bottom of Fig.~\ref{fundamentalidentities}.}
\end{figure}

The definition in terms $\delta$ and $\gamma$ is $(A\odot B)_{mn} = A_{ij}B_{k\ell}\gamma_{ik,m}\delta_{j\ell n}$ which means that we perform a Kronecker product on the row idices and a Hadamard product on the column indices.
\begin{figure}[h!]
\centering
\includegraphics[page=12, width = 0.5\columnwidth]{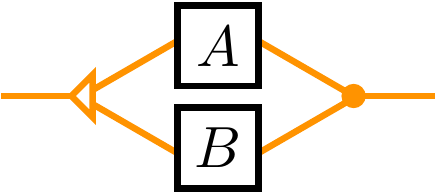}
\caption{In the Katri-Rao product we perform a Kronecker product row-wise and a Hadamard product column-wise (which is why the number of columns of $A$ and $B$ must match).}
\end{figure}
There is also a row-version of the Kathri-Rao product, where the Hadamard product is on the left (i.e. on the rows), and the vectorization on the right (i.e. on the columns). The graphical definition makes it trivial to prove that the column- and row- versions of the Kathri-Rao product are linked by a transpose operation like so: $(A\odot_\mathrm{col} B)^T = A^T\odot_\mathrm{row} B^T$ (see Fig.~\ref{KR1}).

\subsection{Tracy-Singh product}
The Tracy-Singh product \cite{tracy1972new} is a double Kronecker product and it applies to block-matrices: the first is at the level of external blocks and the second is at the inner (in-block) level. Finally, outer and inner indices are vectorized to obtain a matrix: $(A\star B)_{mn} = A_{ijk\ell}B_{pqrs}\gamma_{ipkr,m}\gamma_{jq\ell s,n}$
\begin{figure}[h!]
\centering
\includegraphics[page=13, width = 0.6\columnwidth]{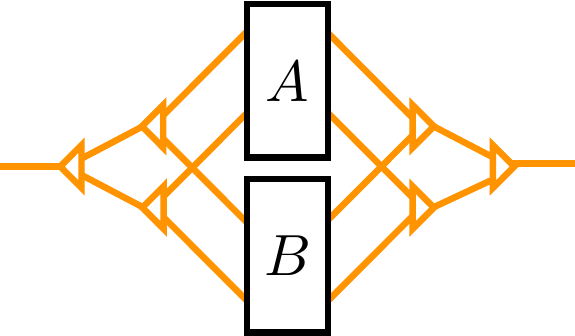}
\caption{In the Tracy-Singh product we first combine the indices of the same level (outer with outer and inner with inner) and only afterward we combine the inner/outer indices with each other. This can be performed in a single step if we are careful with the order.}
\end{figure}

\section{identities}
Thanks to our graphical notation we can easily prove many identities that connect the various products to each other. Several are in the form
\begin{align}
f(g(A,B),g(C,D)) = g(f(A,C),f(B,D))
\end{align}
where $f$ and $g$ can be some combinations of the products that we have seen above, and $A, B, C, D$ are suitably sized matrices or tensors. Such property is called \emph{bysimmetry} (notice that $B$ and $C$ are swapped) and it is due to the fact that $f$ and $g$ are homomorphisms that preserve each other \cite{aczel1948mean}. 

The proofs consist in visual representations of the statement that we want to prove. Note that often just writing the identities in graphical notation is sufficient to realise at a glance that they are true, without the need to dive into lengthy computations. All of the necessary comments are in each figure caption.
 
\begin{figure}[h!]
\centering
\includegraphics[page=14, width = 0.4\columnwidth]{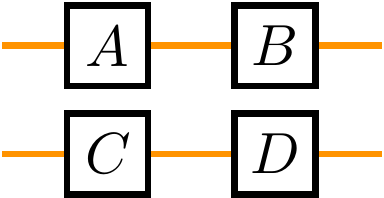}
\caption{\label{TD1}In this figure we prove the following property of the tensor product: $(A B)\otimes(C D) = (A\otimes C)(B\otimes D)$. To convince ourselves that this identity holds, we just have to read the diagram from top to bottom or left to right to obtain the two sides of the identity.}
\end{figure}

\begin{figure}[h!]
\centering
\includegraphics[page=15, width = \columnwidth]{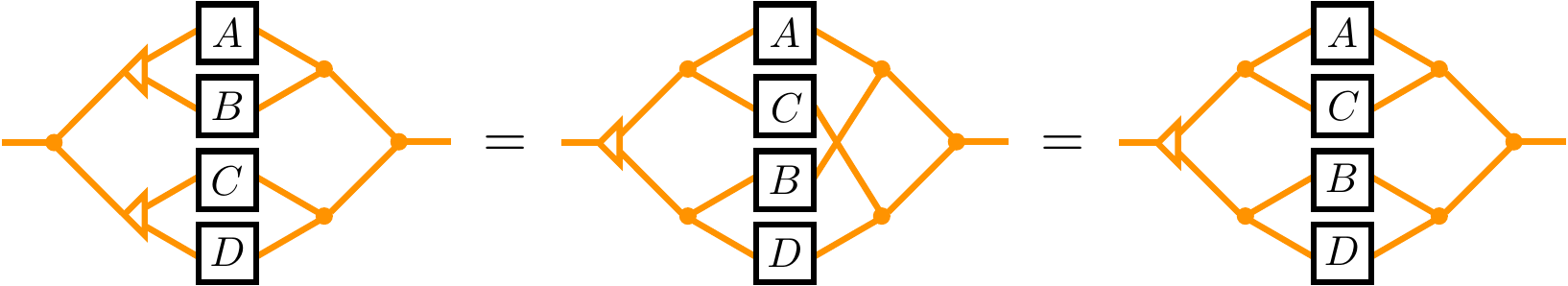}
\caption{\label{KRH2}In this figure we prove the bisymmetry between Kathri-Rao and Hadamard products:  $(A\odot B)\circ(C\odot D) = (A\circ C)\odot (B\circ D)$. To prove this identity we have applied the rule in Fig.~\ref{KVswap} to the wires on the left. Note that all of the wires on the right belong to the same rank-5 Kronecker tensor, so it does not matter where they attach to.}
\end{figure}

\vspace{1mm}
\begin{figure}[h!]
\centering
\includegraphics[page=16, width = \columnwidth]{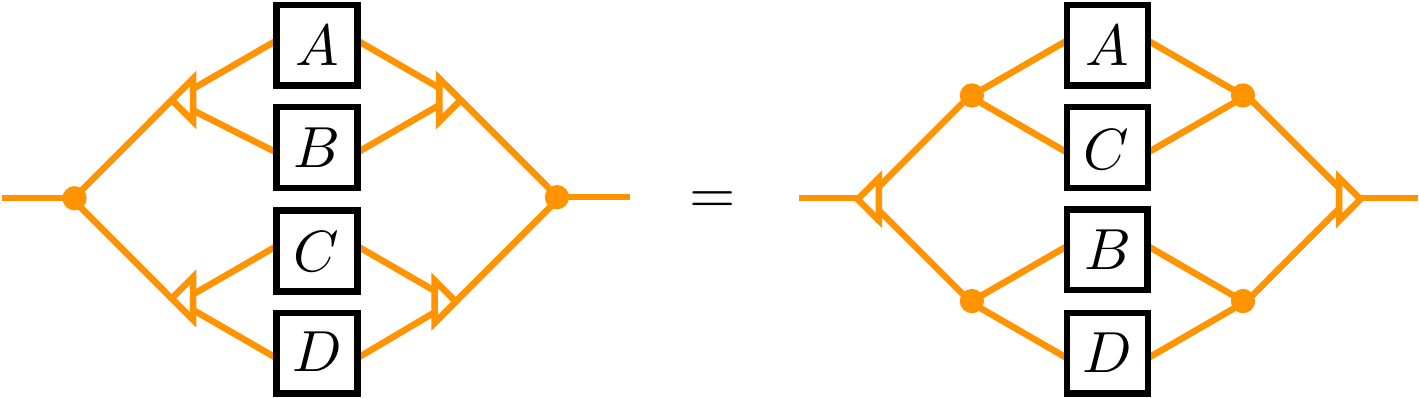}
\caption{\label{KH1}In this figure we prove the bisymmetry between Kronecker and Hadamard products: $(A\otimes B)\circ(C\otimes D) = (A\circ C)\otimes (B\circ D)$. To prove this identity we have applied the rule in Fig.~\ref{KVswap} to the wires on both sides.}
\end{figure}

\begin{figure}[h!]
\centering
\includegraphics[page=17, width = 0.8\columnwidth]{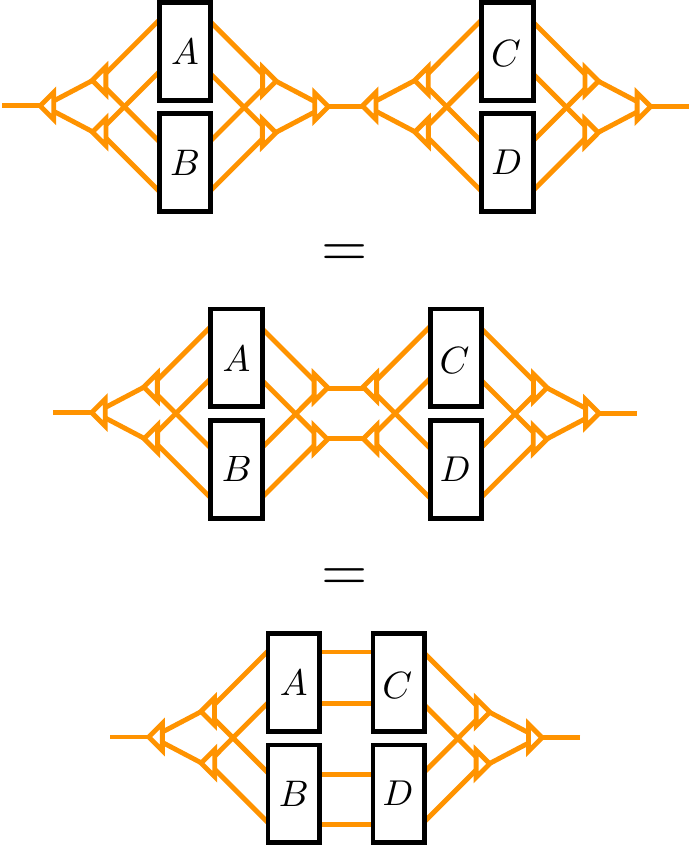}
\caption{\label{TS1}In this figure we prove $(A\star B)(C\star D) = (AC)\star (BD)$. To prove this identity we have applied the rule at the bottom of Fig.~\ref{fundamentalidentities}.}
\end{figure}

We leave as a useful exercise for the reader to prove $A\otimes B=(A\otimes \mathbbm{1})(\mathbbm{1}\otimes B)$, $\mathrm{Tr}(A\otimes B)=\mathrm{Tr}(A)\mathrm{Tr}(B)$ and $(A\otimes B)^T=A^T\otimes B^T$.

\subsection{Identities involving vectorization}
Vectorization (as introduced above through the tensor $\gamma$) is the procedure of turning a high-rank tensor into a column or row vector by stacking its entries in some order. For notational simplicity, when we stack the columns of the matrix $A$ into a column vector we will indicate the operation with $\mathrm{col}(A)$, and when we concatenate the rows of $A$ into a row vector we will indicate it with $\mathrm{row}(A)$. The two are related by $\mathrm{col}(A)^T = \mathrm{row}(A^T)$, as stacking the columns of a matrix and then transposing the resulting vector is the same as transposing the matrix and then concatenating the rows.

The vectorization of an $M\times N$ matrix is a column or row vector with $MN$ elements. Note that we could either truly form a vector that has one single index, or we could reinterpret both indices of $A$ as column (or both as row) indices, in which case we would have a ``block vector'' with the first index indexing $N$ blocks of dimension $M\times 1$  (or $M$ blocks of dimension $1\times N$) and a second index indexing the elements within each block. In any case, given that the vectorization tensor is reversible, conflating vectorization and block-vectorization is just as acceptable as conflating Kronecker and tensor products.
\begin{figure}[h!]
\centering
\includegraphics[page=19, width = 0.6\columnwidth]{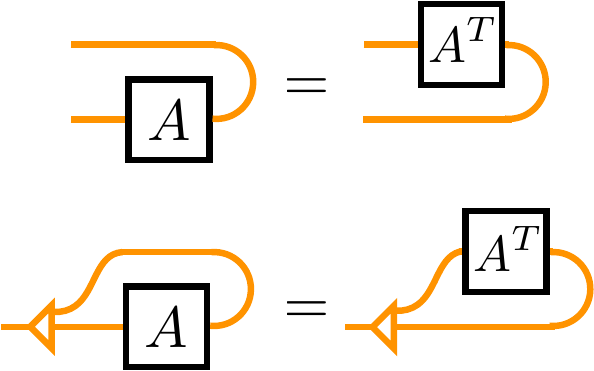}
\caption{\label{vectorization}A block vector (top) and a vectorized matrix (bottom). We denote both with $\mathrm{col}(A)$, as it is usually clear from the context which we are referring to.}
\end{figure}

One can come up with several interesting identities involving vectorization. We begin with an identity involving the vectorization of a product of three matrices and the Kathri-Rao product or the Kronecker product, depending on whether or not the matrix $B$ is diagonal. For the case in which $B$ is not diagonal, we first interpret the vectorization of $B$ as block-vectorization, and then we use the identity at the bottom of Fig.~\ref{fundamentalidentities}. For the case in which $B$ is diagonal we can use the first rule in Fig.~\ref{KTp}.

\begin{align}
\mathrm{col}(ABC) &= (C^T\otimes A)\mathrm{col}(B)\quad [B\ \mathrm{not\ diagonal}]\\
\mathrm{col}(ABC) &= (C^T\odot A)\mathrm{diag}(B)\quad [B\ \mathrm{diagonal}]
\end{align}
Note that in the first equation we use a Kronecker product (not a tensor product).
\begin{figure}[h!]
\centering
\includegraphics[page=1, width = \columnwidth]{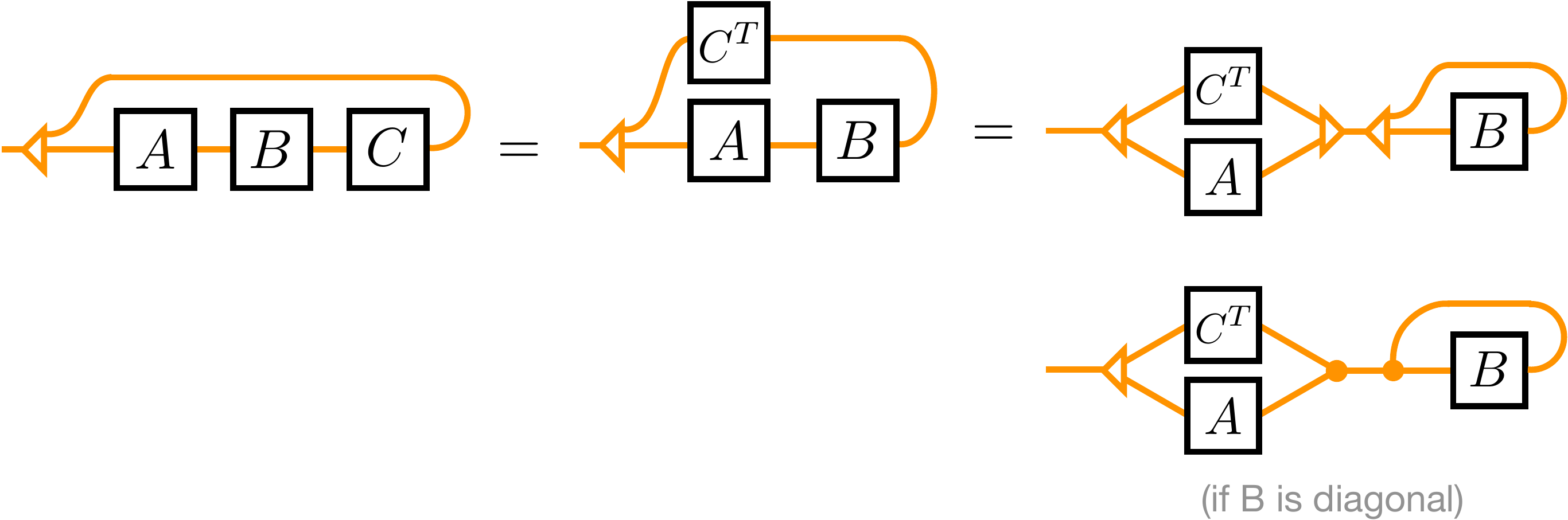}
\caption{\label{vec1}This type of manipulation is found for example in optimization problems \cite{yang2008some}. In this proof we used identities from Fig.~\ref{KTp} and Fig.~\ref{fundamentalidentities}.}
\end{figure}

\section{New identities}
\begin{figure}[h!]
\centering
\includegraphics[page=2, width = 0.9\columnwidth]{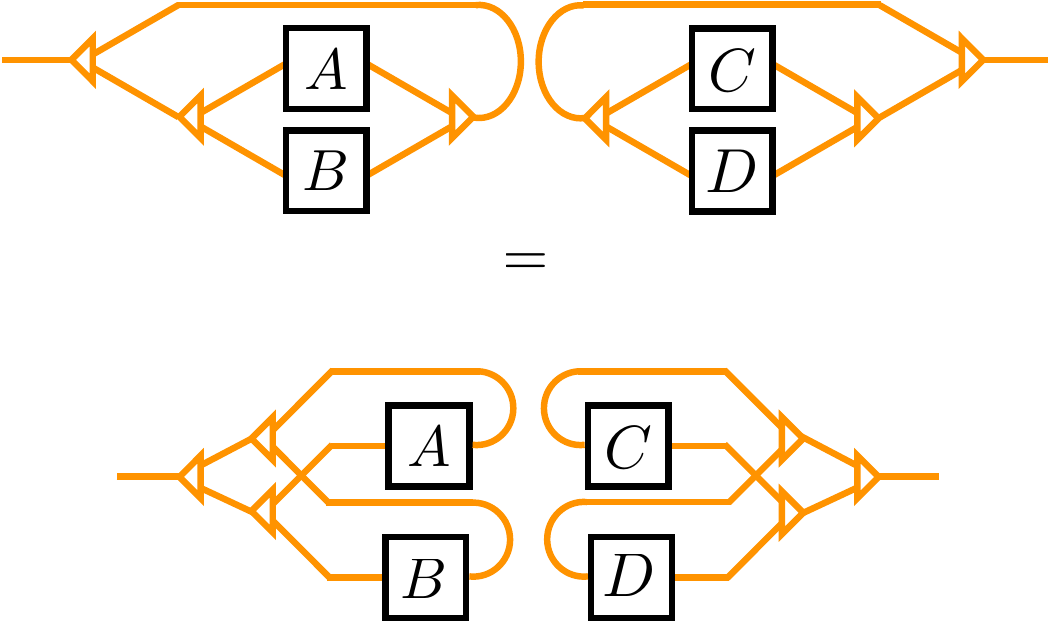}
\caption{\label{TSvec}In this proof of the identity $\mathrm{col}(A\otimes B) \mathrm{row}(C\otimes D) = (\mathrm{col}(A)\mathrm{row}(C))\star(\mathrm{col}(B)\mathrm{row}(D))$ we simply drag the inner triangles outwards and we notice that we recover the definition of the Tracy-Singh product.}
\end{figure}

\begin{figure}[h!]
\centering
\includegraphics[page=3, width = 0.9\columnwidth]{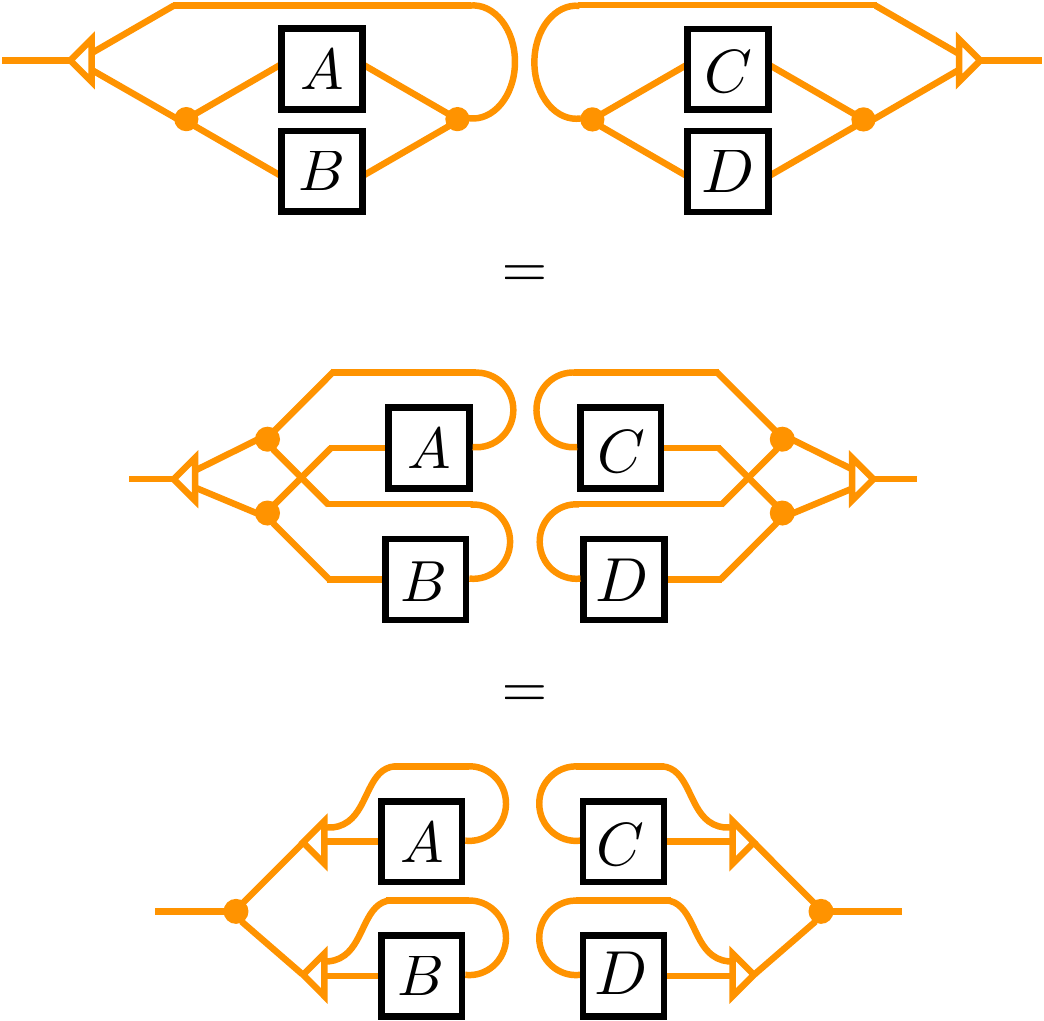}
\caption{\label{Hvec}In this proof of the identity $\mathrm{col}(A\circ B)\mathrm{row}(C\circ D) = (\mathrm{col}(A)\mathrm{row}(C))\circ(\mathrm{col}(B)\mathrm{row}(D))$, we swap the Kronecker and vectorization tensors, which implies twisting the inner wires and leaves us with a neat Hadamard product. Note that the two parts (left and right), if considered in isolation, form two identities on their own: $\mathrm{col}(A\circ B) = \mathrm{col}(A)\circ\mathrm{col}(B)$ and $\mathrm{row}(C\circ D) = \mathrm{row}(C)\circ\mathrm{row}(D)$.}
\end{figure}

To show the flexibility of graphical notation, we now prove in Fig.~\ref{TSvec} and Fig.~\ref{Hvec} two new (to the best of the author's knowledge) identities. The first identity is 
\begin{align}
\mathrm{col}(A\otimes B) \mathrm{row}(C\otimes D) = (\mathrm{col}(A)\mathrm{row}(C))\star(\mathrm{col}(B)\mathrm{row}(D))
\end{align}
which involves vectorization, the Kronecker product and the Tracy-Singh product.

The second identity is
\begin{align}
\mathrm{col}(A\circ B)\mathrm{row}(C\circ D) = (\mathrm{col}(A)\mathrm{row}(C))\circ(\mathrm{col}(B)\mathrm{row}(D))
\end{align}
involving Hadamard product and vectorizations.

Notice that the proof in Fig.~\ref{Hvec} can be separated into two parts (the left and the right parts), which yield the identities
\begin{align}
\mathrm{col}(A\circ B) &= \mathrm{col}(A)\circ\mathrm{col}(B)\\
\mathrm{row}(C\circ D) &= \mathrm{row}(C)\circ\mathrm{row}(D)
\end{align}
where the Hadamard product of two vectors is still the element-wise product. This can be used to write the second identity in yet a different way by reading the last diagram left to right rather than top to bottom.

As you look at the figures, notice how simple it is to observe differences and similarities between proofs: the structure of the diagrams in Fig.~\ref{TSvec} and \ref{Hvec} is almost identical, with the exception that in the second proof, in order to swap the Kronecker and vectorization tensors we need to twist the central wires, which is why we obtain the Hadamard product of distinct vectorized matrices.

\section{Convolutions}
We now turn to the graphical description of convolutions. The mechanisms of convolution and cross-correlation can be nicely translated to graphical calculus notation, and one can notice several interesting features visually. In particular, we will see that convolution is represented just like the products that we have described in the previous sections, and that the (discrete) convolution theorem is a special case of a more general tensor contraction rule that is entirely independent from the tensors being convolved.


\subsection{The convolution tensor}
We now define a general convolution tensor that can implement several different types of convolution. The convolution tensor is a family of $D$-dimensional, rank-3 tensors defined as
\begin{align}
\chi_{ijk}^{(\pm\pm\pm)} = \begin{cases}1\quad \mathrm{if}\ \pm i\pm j\pm k=0\,\mathrm{mod}\,D\\0\quad \mathrm{otherwise}\end{cases}
\end{align}
where the signature $(\pm\pm\pm)$ determines four possibilities (under the $+\leftrightarrow -$ symmetry). In graphical notation, we define the convolution tensor as 
\begin{figure}[h!]
\centering
\includegraphics[page=4, width = 0.4\columnwidth]{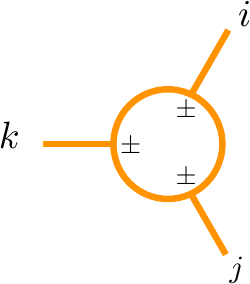}
\caption{\label{convolution}The convolution tensor depends on three signs, which determine which combination of indices yields a nonzero value.}
\end{figure}

Depending on the signature, this tensor implements different convolutions between two vectors $(\mathbf{a}*\mathbf{b})_k^{(\pm\pm\pm)} = \chi_{ijk}^{(\pm\pm\pm)} a_ib_j$, or more explicitly:
\begin{align}
(\mathbf{a}*\mathbf{b})_k^{(\pm\pm\pm)} = \sum_{i=0}^{D-1}a_{\pm i}b_{\pm(\mp k-\mp i)\,\mathrm{mod}\, D}
\end{align}

\begin{figure}[h!]
\centering
\includegraphics[page=6, width = 0.35\columnwidth]{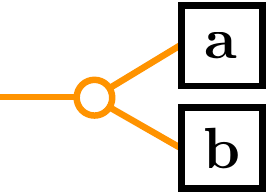}
\caption{\label{convolution} The convolution of two vectors can be expressed as a product.}
\end{figure}

Such definitions include the conventional discrete convolution ($++-$) and the cross-correlation ($+--$):
\begin{align}
(\mathbf{a}*\mathbf{b})_k^{(++-)} &= \sum_{i=0}^{D-1}a_{i}b_{(k- i)\,\mathrm{mod}\, D}\\
(\mathbf{a}*\mathbf{b})_k^{(+--)} &= \sum_{i=0}^{D-1}a_{i}b_{(k + i)\,\mathrm{mod}\, D}
\end{align}

Note that the cross-correlation (usually in its non-circular version) is what is usually referred to as ``convolution'' in the context of convolutional neural networks.

It is straightforward to extend this definition to higher rank convolutions, e.g.~for matrices (where we intend all index algebra to be modulo the dimension of the index), the standard definition is: 
\begin{align}
(A*B)_{mn} = \sum_{i=0}^{D^{x}-1}\sum_{j=0}^{D^{y}-1}A_{i,j}B_{m-i,n-j}
\end{align}
which corresponds to the following graphical diagram where the convolutions have signature $(++-)$:
\begin{figure}[h!]
\centering
\includegraphics[page=7, width = 0.5\columnwidth]{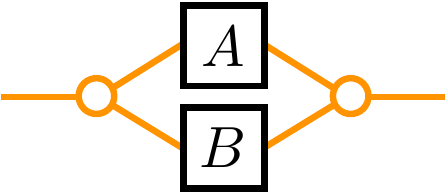}
\caption{\label{convolutionMatrices} The convolution of two matrices $A$ and $B$ can be expressed as a suitable tensor contraction and therefore has the same visual representation as a matrix product. (notice that we could also apply convolutions of different signature to different sets of indices)}
\end{figure}

And one can continue in a similar fashion, for an increasing number of indices.
Interestingly, even if a convolution kernel is not separable (e.g.~ $K_{ij} \neq K_i^xK_j^y$), the ``process'' is, which means that rank-$N$ convolution is computable by contracting two indices at a time.

\subsection{The convolution theorem}
The convolution theorem states that under the Fourier transform, products become convolutions and vice versa:
\begin{align}
\widehat{f(x)g(x)} &= \widehat f(k)*\widehat g(k)
\end{align}

\begin{figure}[h!]
\centering
\includegraphics[page=8, width = 0.9\columnwidth]{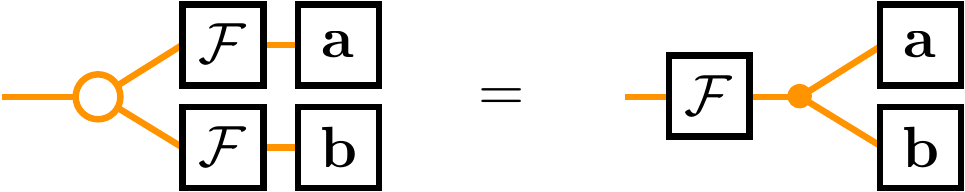}
\caption{\label{convolutionthm} The standard convolution theorem where the convolution tensor has signature $(++-)$}
\end{figure}

If we discretize the functions into $D$-dimensional vectors, the Fourier transform becomes a matrix multiplication by $F_{mn} = \frac{1}{\sqrt{D}}\exp(-\frac{2\pi i}{D}mn)$ and the product becomes the Hadamard (element-wise) product (see Fig.~\ref{convolutionthm}).

As the convolution theorem must hold regardless of the vectors being convolved, it must be a consequence of the interplay between Fourier matrices and the convolution tensor itself: when we contract each index of the convolution tensor by a Fourier matrix, we obtain the Kronecker tensor, which is what we need to implement the Hadamard (entry-wise) product. However, depending on the three signs of the convolution tensor, the convolution theorem takes a slightly different form: indices that have a plus sign are contracted by the Fourier matrix and indices that have a minus sign are contracted by the inverse Fourier matrix (or vice versa, under the $+\leftrightarrow -$ symmetry). See Fig.~\ref{convolution} for an example.
\begin{figure}[h!]
\centering
\includegraphics[page=5, width = 0.9\columnwidth]{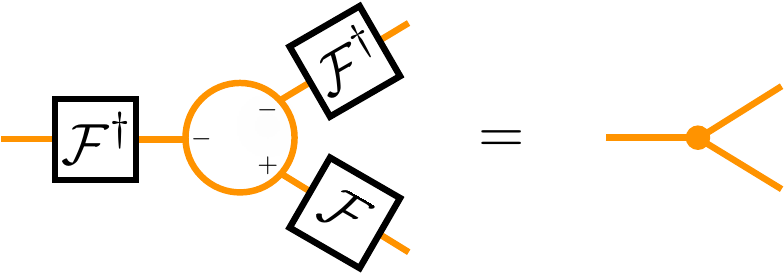}
\caption{\label{convolution} The convolution tensor in this example satisfies the generalized convolution theorem where indices with opposite sign in $\chi_{ijk}^{(\pm\pm\pm)}$ get contracted by opposite Fourier transforms and always yield the Kronecker tensor.}
\end{figure}

\section{Tensor contractions in python}
Python's numerical math library \texttt{numpy} contains a function called ``\texttt{einsum}'' (short for Einstein summation convention) \cite{oliphant2006guide}. Learning how to use \texttt{einsum} can be quite challenging without a good grasp of the rules of tensor contraction. This is where graphical calculus can help.

There are two ways to use \texttt{einsum}: the first is to define all the contraction rules in the form of a string (passed as first argument) that specifies what happens to the indices of the various tensors passed as subsequent arguments (one can pass several tensors, not just one or two). For example, if we want to perform the matrix-matrix multiplication $A_{ij}B_{jk}$, the string would read \texttt{"ij,jk -> ik"}, where we can specify in which order we want the leftover indices. A tensor product $A_{ij}B_{kl}$ would correspond to the string \texttt{"ij,kl"}, a partial trace $A_{iijk}$ to \texttt{"iijk"} and so on:
\begin{lstlisting}
AdotB = einsum("ij,jk->ik", A,B)
AtensorB = einsum("ij,kl", A,B)
tr1A = einsum("iijk", A)
\end{lstlisting}

At the time of writing (March 2019), there are some functionalities that are still missing from \texttt{einsum}, such as the possibility of creating a diagonal matrix from the vector of the diagonal, which would correspond to the string \texttt{"i -> ii"}.

The second way to use \texttt{einsum} does not require a string with the index specifications. Rather, the arguments of \texttt{einsum} alternate between tensors and lists of integer numbers that identify their indices. When an index repeats it is summed over. The matrix multiplication, tensor product and partial trace in the paragraph above would be \begin{lstlisting}
AdotB = einsum(A,[0,1],B,[1,2],[0,2])
AtensorB = einsum(A,[0,1],B,[2,3])
tr1A = einsum(A,[0,0,1,2])
\end{lstlisting}

It is of great help to draw by hand the required network of tensor contractions graphically, then identify the wires by number or by letter (thus wires that contract have a single number or letter) and finally simply copy the formula in the language of \texttt{einsum}. This is a foolproof way of implementing tensor contractions in python.

\section{conclusion}
In this paper we have seen two of the many potential applications of graphical calculus. The message that I wish to convey is that graphical calculus is a powerful companion to the student and to the practitioner of linear algebra. As innumerable topics in science are based on linear algebra, graphical calculus can become an invaluable ally that strengthens our understanding and makes us reach further with less effort. The fact that visual representations are much simpler to grasp and manipulate than conventional mathematical expressions, makes it a wonderful tool also for audiences who lack a knowledge of university-level algebra.

Finally, I would like to point out that these sort of symbolic manipulations lend themselves very well to gamification. I believe that it would be valuable to explore computer and smartphone applications that can teach linear algebra (even rather complex topics) through the gamified manipulation of graphical objects.

\section{acknowledgements}
I thank Jacob Biamonte for several useful conversations and Electra Eleftheriadou for her support and helpful feedback on this manuscript.

\bibliographystyle{unsrt}
\bibliography{tnp}

\end{document}